# Java File Security System (JFSS) Evaluation Using Software Engineering Approaches

Dr. Brijender Kahanwal,*
ASSOCIATE PROFESSOR,
COMPUTER SC. & ENGG. DEPTT.
GGGI, DINARPUR, AMBALA, HARYANA
(INDIA)

Dr. Tejinder Pal Singh,
ASSOCIATE PROFESSOR,
*DEPARTMENT OF APPLIED SCIENCES,*
*RPIIT, BASTARA, KARNAL, HARYANA*
*(INDIA)*

*Abstract* -- A Java File Security System (JFSS) [1] has been developed by us. That is an ecrypted file system. It is developed by us because there are so many file data breaches in the past and current history and they are going to increase day by day as the reports by DataLossDB (Open Security Foundation) organization, a non-profit organization in US so it is. The JFSS is evaluated regarding the two software engineering approaches. One of them is size metric that is Lines of Code (LOC) in the software product development. Another approach is the customer oriented namely User Satisfaction Testing methodology.

Satisfying our customers is an essential element to stay in business in modern world of global competition. We must satisfy and even delight our customers with the value of our software products and services to gain their loyalty and repeat business. Customer satisfaction is therefore a primary goal of process improvement programs as well as quality predictions of our software. With the help of User Satisfaction Index that is calculated for many parameters regarding the customer satisfaction. Customer Satisfaction Surveys are the best way to find the satisfaction level of our product quality.

## I    INTRODUCTION

There is a great deal of confidential information in govt. sector or private sector or other type of organizations about their employees, customers, products, research, and financial status. Now almost information is collected, processed and stored on electronic computers and transmitted across networks to other computers. Should confidential information about a business customers or finances or new product line fall into the hands of a competitor, such a breach of security could lead to lost business, law suits or even bankruptcy of the business. Protecting it is a business requirement, and in many cases also an ethical and legal requirement. For the individual, information security has a significant effect on Privacy, which is viewed very differently in different cultures. In computer systems, all the information is stored traditionally in form of files or databases. File is considered as a basic entity for keeping the information. In UNIX operating system, the concept of file is so important that almost all types of devices are considered as a file. So this problem of securing data or information on computer systems can be identified as the problem of securing files. In the current scenario, securing file data is very significant.

There are various approaches available to ensure file data security. But each one has its own inherent disadvantages, providing them less frequently used. These approaches take the help of cryptography. But all have their own working areas that may be kernel level cryptography or user level cryptography in the operating system. Security approach can be applied at hardware or software levels. We have developed the Java File Security System (JFSS) in the software level and in the user level of the operating system. In this paper we had made the survey for assessing the quality of the software regarding customer's point of view.

Basically the user satisfaction testing is done. One hundred users have been selected and all are computer savvy because they are in the need of the information security and they know the value of information. A naïve user will not be interested in it because he is the new to the computer system. He/ she will take a lot of time to learn the computer basics and then he can help us in this testing. So the sample for the survey is UG engineering students from all branches and the staff members of the institute.

## II    METHODOLOGY

Software complexity is traditionally a direct indicator of software quality and cost. The greater the complexity (by some measure) the more fault prone the soft-ware resulting in higher cost. Code size is one of the easiest things to measure. The code size can be used as a predictor of software characteristics such as effort and ease of maintenance. The line of code (LOC) [17] is a measure for complexity of the software.

Many useful comparisons involve only the order of magnitude of lines of code in a project. Software projects can vary between 1 to 100,000,000 or more lines of code. Using lines of code to compare a 10,000 line project to a 100,000 line project is far more useful than when comparing a 20,000 line project with a 21,000 line project. There are some good points of this measure which are as:

**Scope for Automation of Counting:** Since Line of Code is a physical entity; manual counting effort can be easily eliminated by automating the counting process. Small utilities may be developed for counting the LOC in a program. However, a code counting utility developed for a specific language cannot be used for other languages due to the syntactical and structural differences among languages.

**An Intuitive Metric:** Line of Code serves as an intuitive metric for measuring the size of software because it can be seen and the effect of it can be visualized.

This measure suffers from some fundamental problems which are as follows:

**Lack of Accountability:** Some think it isn't useful to measure the productivity of a project using only results from the coding phase, which usually accounts for only 30% to 35% of the overall effort.

**Lack of Cohesion with Functionality:** Though experiments have repeatedly confirmed that effort is highly correlated with LOC, functionality is less well correlated with LOC. That is, skilled developers may be able to develop the same functionality with far less code, so one program with less LOC may exhibit more functionality than another similar program.

**Adverse Impact on Estimation:** Because of the fact presented under point #1, estimates based on lines of code can adversely go wrong, in all possibility.

**Developer's Experience:** Implementation of a specific logic differs based on the level of experience of the developer. Hence, number of lines of code differs from person to person. An experienced developer may implement certain functionality in fewer lines of code than another developer of relatively less experience does, though they use the same language.

**Difference in Languages:** Consider two applications that provide the same functionality (screens, reports, databases). One of the applications is written in C++ and the other application written in a language like COBOL. The number of function points would be exactly the same, but aspects of the application would be different. The lines of code needed to develop the application would certainly not be the same.

**Advent of GUI Tools:** With the advent of GUI-based programming languages and tools such as Visual Basic, programmers can write relatively little code and achieve high levels of functionality. For example, instead of writing a program to create a window and draw a button, a user with a GUI tool can use drag-and-drop and other mouse operations to place components on a workspace. Code that is automatically generated by a GUI tool is not usually taken into consideration when using LOC methods of measurement.

**Problems with Multiple Languages:** In today's software scenario, software is often developed in more than one language. Very often, a number of languages are employed depending on the complexity and requirements. Tracking and reporting of productivity and defect rates poses a serious problem in this case since defects cannot be attributed to a particular language subsequent to integration of the system. Function Point stands out to be the best measure of size in this case.

**Lack of Counting Standards:** There is no standard definition of what a line of code is. Do comments count? Are data declarations included? What happens if a statement extends over several lines?

**Psychology:** A programmer whose productivity is being measured in lines of code will have an incentive to write unnecessarily verbose code. The more management is focusing on lines of code, the more incentive the programmer has to expand his code with unneeded complexity.

Software complexity is widely regarded as an important determinant of software maintenance costs. Increased software complexity means that maintenance and enhancement projects will take longer, will cost more, and will result in more errors. What is more, the software complexity of a given system is one of the main long-term legacies of whatever tools and techniques were employed in its initial development.

**User Satisfaction Testing** [15]

User satisfaction test is a test which is conducted to assess the level of comfort users experience on all dimensions of quality. The survey is done with the help of engineering students and their teaching staff. In order to test the satisfaction level of users, there are two tests are proposed:

- o Usability testing
- o User satisfaction testing  Definition of Usability by ISO is as follows:

―Usability is the effectivness, efficiency and satisfaction with which a specified set of users can achieve a specified set of tasks in a particular environment.

The usability testing measures ease of use and comfort that the users have while using the software. The poor usability directly affects the success of the software because the users find it difficult to learn and difficult to operate resulting in non-usage and eventually the software is termed as unsuccessful.

The usabilty of the software must be looked into right from the beginning, and development should be done keeping user environment and their capability in mind. Usability tests are conducted using use cases.

User satisfaction test is another test that measures, by some attributes of usability, functions, features and cost. Usabilty can be measured by defining measurable goals that confirm the high level of usabilty.

User satisfaction test is conducted asking users to respond to the parameters listed. The Table summarizes the responses of 100 users on seven parameters of satisfaction, and on overall Java File Security System (JFSS).

**User Satisfaction Index (USI) = (Actual Score / Total Highest Score) × 100**

III RESULTS

Here we have applied two software engineering approaches namely User Satisfaction Testing and Lines of Code size metric for complexity of the software. Users behavior is evaluated regarding the file security system. The product JFSS is handovered to the one hundrad of users of diverse qualifications. But all of them are literate one and have some hand on the computer system because the software is for the users who want to secure their data. The naïve users have no use of the product. In the next approach the number of coding lines are to be compared with the existing file security systems. Which shows the compexity of the security system.

**User Satisfaction Index (USI) = (Actual Score / Total Highest Score) × 100**

Here Actual Score for the parameter is counted as a cumulative weighted score of 100 responses is the sum of score multilplied by the number of responses.

**Actual Score (Parameter) = Sum (Score on Satisfaction × Number of responses)** and the **Total Highest Score** is common for all that is 5*100 = 500.

User Satisfaction Index (USI) of all the individual parameters:

1. *Ease of use:*
   Actual Score = (10*1 + 10*2 + 5*3 + 25*4 + 50*5) = 395, USI = (395/500) *100= 79%
2. *Ease of learning:*
   Actual Score = (15*1 +20*2 + 65*5 = 380, USI = (380/500) *100 = 76%
3. *Scope of coverage:*
   Actual Score = 8*1 + 2*2 + 55*4 + 35*5 = 407, USI = (407/500) *100 = 81.4%
4. *SRS coverage:*
   Actual Score = 6*1 + 5*2 + 16*4 + 73*5 = 445, USI = (445/500) *100 = 89%
5. *Performance:*
   Actual Score = 45*3 + 30*4 + 25*5 = 380, USI = (380/500) *100 = 76%
6. *Security:*
   Actual Score = 100*5 =500, USI = (500/500) *100 = 100%
7. *User Interface Quality:*
   Actual Score = 35*4 +65*5 = 465, USI = (465/500) *100 = 93%

Arithmetic Mean (AM) of all the seven parameter's USI is such as:

AM (USI) = (79 + 76 + 81.4 + 89 + 76 + 100 + 93) / 7 = 84.91%

The results of this survey show that the User Satisfaction Index (USI) is 84.91% in overall or in average calculated hare. The file security system is good in use. The user are feeling comfort in the product.

The parameter **Ease of Use** is the direct concern of the user's comfort ability in using the product JFSS. The User Satisfaction Index for this parameter that is 79% which shows that there are only 21% persons which are telling that they are not feeling better in using the software. We have analyzed that this percentage has the users which are not daily users of the computers and they are the persons who comes in touch of the computers hardly in a week.

The second parameter **Ease of Learning**, there is the percentage of User Satisfaction Index (USI) that is 76. It means that the people have shown the interest in learning the new product. And only few of them are not interested. We have provided the learning material to the users that were the handouts of the screen shots and all was fully explained on the projector.

The 3$^{rd}$ parameter is **Scope of coverage** which is about the working area of the software. As we know that the Java File Security System (JFSS) make secure to all types of files whatever it is, like .exe, .pdf, .jpg, etc. But there are all types of users which have applied on the few types of files. So they cannot evaluate the complete scope coverage area of the software.

Table 3.1: The responses of the 100 users regarding different parameters of user satisfaction for the evaluation of Java File Security System (JFSS).

| Sr. No. | Parameters | Score on Satisfaction | | | | | USI % AGE |
|---|---|---|---|---|---|---|---|
| | | 1 Unacceptable | 2 Low | 3 Medium | 4 High | 5 Highly Acceptable | |
| 1 | Ease of use | 10 | 10 | 5 | 25 | 50 | 79 |
| 2 | Ease of learning | 15 | 20 | --- | --- | 65 | 76 |
| 3 | Scope of coverage | 8 | 2 | --- | 55 | 35 | 81.4 |
| 4 | SRS coverage | 6 | 5 | --- | 16 | 73 | 89 |
| 5 | Performance | --- | --- | 45 | 30 | 25 | 76 |
| 6 | Security | --- | --- | --- | --- | 100 | 100 |
| 7 | User Interface Quality | --- | --- | --- | 35 | 65 | 93 |
| Arithmetic Mean (AM) of all seven User Satisfaction Indexes: | | | | | | | 84.91 |

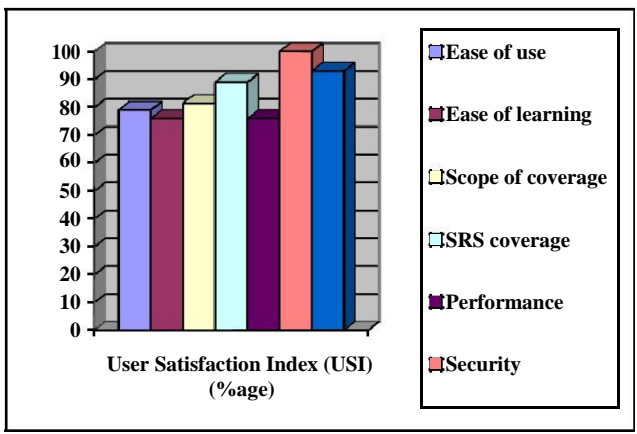

Figure 3.1: A Column chart for the user satisfaction index regarding parameters of user satisfaction testing

The 4$^{th}$ parameter namely **SRS coverage**, the SRS document is provided to the users to check out the software according to that. The User Satisfaction Index (USI) for the same is 89%, which is better one means it is the development according to the SRS document.

The next parameter is **Performance**, which is one of the crucial objectives. The users have observed that the software is working well. But we know that performance is not better as compared to other file security systems, because the Java language is portable and not so good in the performance as compared to the other ones. The user's point of view is good as the USI that is 76% that is not bad one.

The 6$^{th}$ parameter is **Security** which is the main objective of the software to be developed as there is the theft cases are going to increase. The users have tried the software well and say it is good file security system. It can encrypt all type of files. It is appreciated by all the users that the software is fully secure one. The User Satisfaction Index for that is 100% which is the highest one and the maximum one.

The next and the last parameter is **User Interface Quality** which tells the quality of the interface for the users to use the software. User Satisfaction Index (USI) value for this parameter is 93%. It means that the user interaction with the screen shots of the software is well. Interfacing is the main area for the interaction of the users.

*Complexity Perspective*

The complexity of modern file systems is going to increase drastically day by day, and it keeps increasing. The CFS [2] has near about 5200 lines of code (LOC). The TCFS [3] has near about 15000 lines of code (LOC).There are 8000 lines of code (LOC) in the ext2 [7] file system in the Linux kernel, ext3 [5] doubles this, and ext4 [11] doubles it again. Similar is in the case of EncFS [4] which starts from the first version EncFS 0.x with 3700 lines of code (LOC), in the next version EncFS 1.0.x, there are 8700 lines of code (LOC) and in the next EncFS 1.1.x, there are 16200 lines of code (LOC). ReiserFS [6] [14] is also a huge file system in which there are 27500 lines of code in the version ReiserFS3 and inthe version ResierFS4 there are almost 31000 lines of code (LOC). The zfs [13] file system has 71312 lines of code (LOC) in its implementation. A working development version of B+ tree file system (btrfs [12]) already has over 52,000 LOC, XFS [8] is over 77,000 LOC, and other network-based file systems easily exceed 100,000 LOC. As for as the Java file Security System (JFSS) [1] [9] [10] is concerned that has nearby 4000 lines of code (LOC) for the specialization of encryption. The amount of functionality that is provided by the modern file systems that is very high. It causes the less performance, energy efficiency and reliability etc.

The growth of complexity is mainly caused by the increasing functionality which is integrated in a file system. There are so many features which are supported by modern file systems like journaling, B-tree-based search for objects, flexible data extents, access control lists (ACLs), extended attributes, encryption, checksumming, etc.

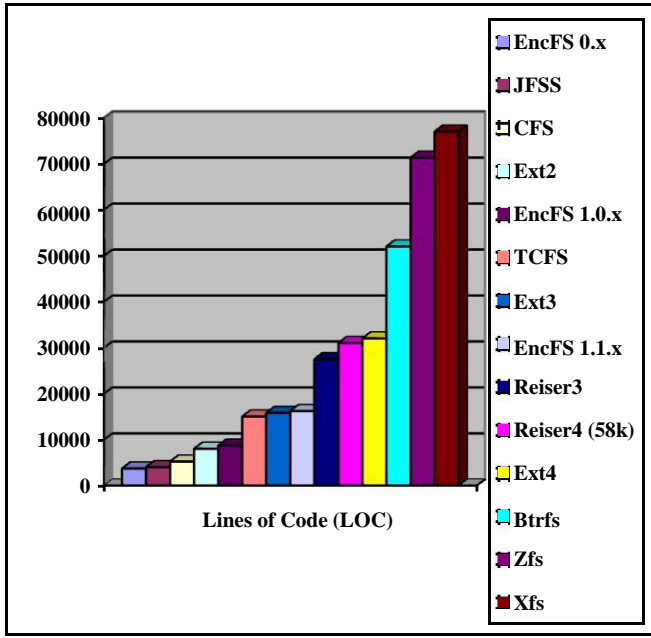

Figure 3.2: The column chart for the file systems according to their implementation size in Lines of Code (LOC)

Some file systems allows writing plug-ins for them like ReiserFS. The large file systems like zfs [13] and btrfs [12] integrate complex storage pool management and deduplication.

The diverse numbers of features are supported by modern file systems to satisfy the end user's appetite of the specific one. On the basis of the specific situation, various characteristics are required from the file system. In some of the cases, reliability is the most important factor, security is essential, servers require high performance, and in mobile systems (Laptops, PDA, Tablets etc.) energy efficiency is an important factor. The developers are going for the integrated file systems which have all the features or the functionalities and as a result the users are getting not only the functionality they require, but also other functionalities those are required by the other users. The file systems are providing so many options to the user that they are in confusion which one is to be used by them.

The file systems with so many functionalities are difficult to develop, maintain and support from the developers point of view. To integrate a new feature to the existing file system takes a lot of time because there is a need to ensure that new functionality interoperates correctly with all other features that are already implemented in the file system. Consequently, the amount of effort grows exponentially for adding the new feature. The number of different code paths in a large file system is huge, leading to an exponential number of states to explore, which considerably complicates debugging and performance analysis. It is difficult to understand the complex file system before fixing bugs or changing file system behavior.

The modern file systems provide so many functionalities to the users. There is no need to use all of these for the users at once. In fact, only minimal file system functionality is sufficient in many cases for the users. Actually, the users are not aware about the basic features of the modern file systems. They have no knowledge about the hard links, soft links, or Access Control Lists (ACLs). There is no use of deep directories. Most of the files often reside in one flat directory. So there is no use of development of all the functionalities in one file system. The file systems should be of specific nature as for as functionality is concerned.

Specific nature's file system should be provided to the needy user. That will be advantageous in all the terms like performance, complexity, cost, lines of code, effort for developing and maintaining a specialized file system.

## IV CONCLUSION

In this article, we have presented the user satisfaction testing results and the complexity of the lines of code within the modern file systems.

In the first evaluation the Java File Security System (JFSS) is evaluated among the hundred users and the complete results show that the JFSS system is good at all and user's are interested in use of it and it provides better security.

In the second evaluation the complexity perspective is discussed hare. As we will increase the functionalities of the file systems then it will lead to the complex system and that will be difficult to maintain, develop and even to understand by the users. The discussion in this article shows that the users are not aware about the functionalities of the file systems. We the researchers are losing the performance of the systems.

At last the results show that JFSS is good in the terms of security of the files. The JFSS is using the main functionalities of the underlying file system.

## ACKNOWLEDGEMENT

We would like to thank all the human beings who were involved in this survey. Special thanks to my old B. Tech. Student, Mr. Kanisk Dua for his great coordination in completing this difficult task.